
\magnification=1200
\hyphenpenalty=2000
\tolerance=10000
\hsize 14.5truecm
\hoffset 1.truecm
\openup 5pt
\baselineskip=18truept
\font\titl=cmbx12



\def\Tg{T_{\gamma}}
\def\({\left(}
\def\){\right)}

\def\kes{\kappa_{es}}

\def\ref{\par\noindent\hangindent 20pt}
\def\refig{\par\noindent\hangindent 15pt}
\def\mincir{\raise -2.truept\hbox{\rlap{\hbox{$\sim$}}\raise5.truept
\hbox{$<$}\ }}
\def\magcir{\raise -4.truept\hbox{\rlap{\hbox{$\sim$}}\raise5.truept
\hbox{$>$}\ }}
\def\rho{\varrho}

\null
\vskip 1.2truecm
\centerline{\titl SPHERICAL ACCRETION ONTO NEUTRON STARS}
\smallskip
\centerline{\titl REVISITED: ARE HOT SOLUTIONS POSSIBLE ?}
\vskip 1.5truecm
\centerline{R.~Turolla $^1$, L.~Zampieri $^2$, M.~Colpi $^3$, A.~Treves $^2$}
\vskip 1.truecm
\centerline{$^1$ Department of Physics, University of Padova}
\centerline{Via Marzolo 8, 35131 Padova, Italy}
\bigskip
\centerline{$^2$ International School for Advanced Studies, Trieste}
\centerline{Via Beirut 2--4, 34014 Miramare--Trieste, Italy}
\bigskip
\centerline{$^3$ Department of Physics, University of Milano}
\centerline{Via Celoria 16, 20133 Milano, Italy}
\bigskip

\beginsection ABSTRACT

Stationary, spherical accretion onto an unmagnetized neutron star is here
reconsidered on the wake of the seminal paper by Zel'dovich \& Shakura (1969).
It is found that new ``hot'' solutions may
exist for a wide range of luminosities. These solutions are characterized by
a high temperature, $10^{9}\div 10^{11}$ K, and arise from a stationary
equilibrium model where the dominant radiative mechanisms are multiple
Compton scattering and bremsstrahlung emission.
For low luminosities, $\mincir 10^{-2} \ L_{E}$, only the ``cold''
(\`a la Zel'dovich and Shakura)
solution is present.

\bigskip
\noindent
{\it Subject headings:\/} accretion, accretion disks \ -- \ stars: neutron
\ -- \ X--rays: stars

\bigskip\bigskip\bigskip
\centerline{Accepted for publication in ApJ Letters}
\bigskip\bigskip
\centerline{Ref. SISSA 21/94/A (February 94)}

\vfill\eject

\beginsection I. INTRODUCTION

Even before the observational evidence that Galactic X--ray sources are
mostly binary systems containing an accreting neutron star,
Zel'dovich \& Shakura (1969, ZS in the following) studied in some detail
the spectrum of radiation produced by stationary, spherical accretion onto an
unmagnetized neutron star and compared their results with the (poor) data
available at the time for Sco X--1. The pioneering paper of ZS shows that the
resulting
spectrum depends on two parameters, the accretion rate (luminosity)
and the penetration length of the accreting ions in the outermost neutron star
layers. The outcome can be described, in essence, as a black body with
a high energy tail due to the Compton heating of thermal photons in the hot,
external part of the atmosphere surrounding the neutron star.

ZS's analytical work was pushed further by Alme \& Wilson (1973), making
use of numerical methods. Shapiro \& Salpeter (1973) adopted essentially
ZS solution at the inner boundary and explored the possibility that in the
surrounding region a shock is formed, which may modify the
resulting spectrum. While models of spherical accretion evolved
substantially since then, considering, for instance, the role of
nuclear reactions induced in the crust by the bombardment of accreting ions
(see e.g. Bildsten, Salpeter \& Wasserman 1992),
the basic picture proposed by ZS has been maintained.

In this letter we reconsider the issue, adopting a set of equations for the
neutron star atmosphere which
essentially coincides with those of ZS. We show that ZS's solution
for the temperature profile is not unique: for large
enough accretion rates, another solution, at considerably higher
temperatures, may exist. The presence of these solutions {\it per se\/},
and the transition between ``hot'' and ``cold'' solutions can have
interesting astrophysical consequences.

\beginsection II. THE MODEL

In order to explore the possible existence of high temperature solutions,
we use a very simplified model which follows closely the original work
of Zel'dovich \& Shakura. In particular we assume that the accreting
flow impinges onto a spherically--symmetric, static atmosphere which
surrounds
the neutron star, and is decelerated as it penetrates into the atmosphere.
The kinetic energy released by the incident protons goes mainly into
electron thermal
energy and is then re--emitted
as free--free radiation. The details of the
flow braking fall within the domain of plasma physics and are still far
from a thorough understanding. For this reason, following ZS, we treat
the total column
density of the atmosphere required to stop the incoming beam, $y_0$, as a
free parameter and the column density

$$ y = \int_R^{\infty} \rho\,  dR \eqno(1) $$
is used as the independent coordinate in place of the radial distance $R$;
here $\rho$ is the matter density.

The heat injected by the infalling protons per unit time and mass
in the atmosphere is assumed to be constant and is
related to the total luminosity observed at infinity by
$$W = {{L_\infty}\over {4\pi \int_0^{y_0} R^2 dy}}\, . \eqno(2)$$
In the inner region where $y>y_0$, $W = 0$;
in all our models $5<y_0<20$ $\rm $. If the flow velocity $v$ exceeds
the electron thermal velocity, $v_{th}$,
these values of the proton penetration length are appropriate
to describe the stopping of the incoming proton beam in a hydrogen
plasma where only Coulomb interactions take place (e.g. Alme and Wilson 1973).
On the contrary, if $v \mincir v_{th}$, the proton stopping through
repeated Coulomb scatterings is less effective and other collective processes
(e.g. plasma oscillations) need to be considered to keep $y_0$ within
this interval.

The transfer of radiation in the atmosphere is governed by the equations
for the radiative luminosity $L$ and the radiation energy density $U$
which, in spherical symmetry and using the Eddington approximation,
can be written as
$$\eqalignno{
 &{{dL}\over{dy}} = - 4\pi R^2 W\, & (3)\cr
 &{1\over 3}{{dU}\over{dy}} = {\kappa}_1{L\over{4\pi R^2 c}} \, .& (4)\cr}$$
The only radiative processes taken into account are scattering and
bremsstrahlung, so that the flux mean opacity can be conveniently expressed as
$${\kappa}_1 = \kes + 6.4\times 10^{22}\rho T^{-7/2} \ {\rm cm^2 \,\,
g^{-1}}\, $$
where $\kes = 0.4 \,\, {\rm { cm^2 \,\,g^{-1}}} \, $ and $T$ is the gas
temperature.
Since, as numerical models show, the atmosphere does not expand considerably
even for high temperature solutions, the Eddington approximation is reasonable.
The appropriate boundary condition for the radiation
field at the outer edge of the non--illuminated medium is
$U = L_\infty/2\pi R^2 c$.
The inner boundary condition, at $y_{in} \gg y_0$, is fixed by the
requirement that all the observed radiative
flux must be generated within the atmosphere, that is, $L = 0$.
This is the same condition used by ZS and is appropriate if the
atmosphere is effectively thick close to $y_{in}$.

The runs of pressure, $P$, and temperature
are obtained from the hydrostatic balance and radiative energy equilibrium
$$\eqalignno{
 & {{dP}\over {dy}} = {{GM_*}\over{R^2}} & (5)\cr
 & {W\over c} = {\kappa}_P\left( aT^4 - U\right) + 4\kes U{{KT}\over
{m_ec^2}}\left(1 - {{\Tg}\over T}\right)\, , & (6)\cr}$$
where $M_*$ is the mass of the neutron star.
The matter density $\rho$ is calculated from the perfect gas equation of
state assuming that the material is completely ionized hydrogen.
Since we study the general properties of spherical
accretion onto neutron stars
for luminosities below the Eddington limit,
in equation (5) we neglected the radiative force and the
ram pressure exerted by the incoming protons, which is typically two orders
of magnitude below the gravitational term.
In equation (6) ${\kappa}_P$ is the Planck mean opacity and $\Tg$ is
the radiation temperature which is defined
as the mean photon energy.
In general $\Tg$ can be computed only solving the full frequency--dependent
transfer problem and will depend on $y$.
In ZS, $\Tg$ was taken equal to $[U(y)/a]^{1/4}$, which is
appropriate in LTE. Being interested also in
solutions in which multiple Compton scattering becomes important, we derive
the radiation temperature from the equation (see Wandel, Yahil \& Milgrom
1984; Park \& Ostriker 1989)

$$ {y\over \Tg}{{d\Tg}\over {dy}} = 2 Y_c \left( {\Tg\over T}
   - 1 \right) \eqno(7)$$
where $Y_c = (4KT/m_e c^2)\max (\tau_{es},{\tau_{es}}^2)$ is the
Comptonization parameter and $\tau_{es} = \kes y$.
Use of equation (7) requires some care since it is meant to describe the
variation of the radiation temperature when multiple Compton scattering
is the dominant mechanism to exchange energy between photons and electrons.
Equation (7) therefore does not apply if either $\tau_{es}<1$ or true
emission--absorption are important. On the other hand, equation (7) gives
the correct limit for low optical depth ($Y_c\ll 1$ implying $\Tg = const$),
so that one can extend the validity of equation (7) to all regimes,
provided that no physical significance is attached to $\Tg$ where the
effective optical depth $\tau_{eff}>1$.

{}From equation (1), it follows immediately that

$${{d R}\over{d y}} = -{1\over\rho}.\eqno(8)$$
The boundary conditions for equations (5) and (8) are $P = 0$
at $y =0$ and $R = R_*$ at $y = y_{in}$.
We notice that a boundary condition for $\Tg$ must be also imposed
because the radiation temperature obeys a
differential equation; models were obtained specifying a value of $\Tg$ at
$y=0$.
The solution to equations (3)--(8) provides the variation of $L$,
$U$, $P$, $T$, $T_\gamma$ and $R$ as functions of the column density $y$ and
is found numerically; all models refer
to a neutron star of $R_* = 10$ km and $M_* = 1\,\, M_{\odot}\, $.

We find that two distinct kinds of
solutions, ``hot'' and ``cold'', always exist for any $y_0$
provided that the luminosity exceeds a certain limit, which depends on $y_0$.
The thermal
properties of the atmosphere are illustrated in figures 1 and 2, where the
run of $T$ versus column density is shown for different
luminosities in the case $y_0=20$.
The ``cold'' solutions of figure 1
are just those already found by ZS and are obtained setting $\Tg(y=0) =
[L_{\infty}/(4\pi R_*^2\sigma)]^{1/4}$. The ``hot'' solutions of figure 2
exist for values of $\Tg$
satisfying the condition $\Tg > T_{crit}(L_{\infty},y_0)$; here $\Tg(y=0) =
2 \times 10^9$ K. Temperature is
close to $\Tg$ in the outer region ($y\mincir 23 \ {\rm g \, cm}^{-2}$), while
in the dense layers
close to $y_{in}$ LTE is attained at $T\sim 10^7$ K. The
temperature profile in the hot region resembles that one
of static atmospheres around X--ray bursting neutron stars
(see e.g. London, Taam \& Howard 1986) where the same radiative processes
dominate.
Cold, thermal photons
do not propagate outwards because $L=0$ for $y_0<y<y_{in}$, so the hot and
the cold zone are thermally decoupled, at least radiatively. On the
other hand, we have checked
that imposing either an ingoing or an outgoing flux at $y_{in}$ does not
alter our picture significantly if $|L(y_{in})|/L_\infty\mincir 0.1$.

The presence of two possible regimes has a simple interpretation in terms of
the relative efficiency of the two radiative processes we have considered,
Compton scattering and free--free emission--absorption.
The static atmosphere must, in fact, radiate a given luminosity and, for doing
that in the scenario we are proposing, there are two ways. A first
possibility is that a lot of soft bremsstrahlung photons are produced in a
low--temperature, dense medium in which the effective depth is large. This
gives rise to a spectrum which is essentially blackbody
and corresponds to the ``cold'' solution. Comptonization
is never dominant because temperature is low and the scattering depth is
not large enough to make $Y_c>1$.
The ``hot'' solutions represent the opposite case, in which much less
energy is generated through bremsstrahlung emission in a low--density,
hot plasma far from LTE. Comptonization, however, is now so efficient
that matter and radiation temperatures are everywhere very close and the same
energy output can be obtained.

It is possible to get an insight on the
existence of high--temperature solutions and to give an estimate of the
limiting value $T_{crit}$ by means of
simple analytical considerations using, for the sake of simplicity,
a plane--parallel geometry for the atmosphere.
For $y<y_0$ we can safely neglect free--free absorption in equation (4) and
we get the expressions for $L$, $U$ and $\rho$ as functions of $y$ and $T$:

$$ L = L_{\infty}{{y_0 - y}\over y_0}  \eqno(9) $$
$$ U = {W\over c} \left[ 2y_0 + 3k_{es}y \left(y_0 -{y\over 2}\right)\right]
   \eqno(10) $$
$$\rho =  D {y\over T}  \eqno(11) $$
where $D = GM_* m_p/2k R_*^2 = 8.1 \times 10^5$, in c.g.s. units
for $R_* = 10^6 \, {\rm cm}$ and $M_* = 1 \,M_{\odot}$. Neglecting the term
$\kappa_P U$, the energy equation becomes a cubic equation in $x = T^{1/2}$,
that can be studied analytically for given values of $y_0$ and $L_{\infty}$
and treating $T_\gamma$ as a free parameter. Equation (6) can be cast
into the form
$$ x^3+p\,x+q=0  \eqno(12) $$
with
$$p=-\left( T_{\gamma} + A {l_{\infty}\over U} \right)
\sim -T_{\gamma}$$
$$q=B {y\over {U(y)}}\, ,$$
where $l_{\infty}$ is the total luminosity observed at infinity in units of
the Eddington luminosity; $A = 6.3 \times 10^{22}$ and $B = 5.1 \times
10^{25}$,
again in c.g.s. units.
The approximated expression for $p$ holds only for $\Tg >> 10^8$K,
while, in order to make the analytical treatment
affordable, in the inner part of the atmosphere
$q$ is set approximately equal to its maximum value, $q_{max} = q(y_0)$.
Once $y_0$ and $l_\infty$ are fixed, equation (12) has one or three
real roots, according to the sign of the discriminant, and
it is easy to prove that all the roots are real only if
$$\Tg \ge {{2.5 \times 10^8}\over
  {\left(1 + {3\over 4}k_{es}y_0 \right)^{2/3} l_{\infty}^{2/3}}} \, .
  \eqno(13) $$
It can also be shown that if just one root is present it is $T\magcir \Tg$,
while, when condition (13) is satisfied,
the three roots have magnitudes $T\magcir \Tg$, $T<<\Tg$ and $T\mincir \Tg$,
respectively. The solution $T\magcir \Tg$ is unacceptable
since $T>\Tg$ will produce a negative radiation temperature gradient
(see eq. [7]) and also $T<<\Tg$ must be discarded because it
is inconsistent with our starting assumption that absorption could
be neglected because the plasma is very hot. Finally, the root
$T \mincir \Tg$ is the ``hot'' solution, which
exists only when the radiation temperature exceeds the limit given by (13),
which represents the analytical estimate of $T_{crit}$.

In figure 3 we plot the mean energy
of the outgoing photons as a function of the
total luminosity for $y_0 = 20$. For ``hot'' solutions only the lower
bound $T_{crit}$ (dashed line) is shown. As can be seen, while the
mean photon energy of the ``cold'' solutions (crosses) monotonically increases
with the luminosity, the lower bound $T_{crit}$ for ``hot'' solutions
shows the opposite behaviour.
Moreover the numerical analysis indicates that high temperature solutions may
exist only for high
enough values of $l_\infty$ and that the critical luminosity , $l_{cr}$,
under which no ``hot'' solutions exist depends on $y_0$: for $y_0=5$,
$l_{cr} = 2 \times 10^{-2}$, while for $y_0=20$, $l_{cr} = 6 \times 10^{-3}$.

The anticorrelation between $T_{crit}$ and $l_{\infty}$,
and the lack of ``hot'' solutions at low luminosities can be
explained as follows.
As $l_{\infty}$ decreases, $W$ and $U$ start to decrease
(see eqs. [2] and [10]) and, since Compton heating (which is dominant over
cooling) becomes progressively smaller in magnitude,
the free--free emissivity ($\kappa_P aT^4 \propto \rho T^{1/2}
\propto P/T^{1/2}$)
must also decrease for the energy equation to be satisfied.
This can be achieved only by an
increase of temperature (and a decrease of density, see eq. [11])
because the pressure profile is nearly independent on the thermal properties
of the gas.
The lack of ``hot'' solutions at small luminosities is due
to the fact that, when density is very low, the envelope becomes photon starved
and even a strong Comptonization is unable to produce the required
luminosity.

The existence of both ``cold''
and ``hot'' solutions for the same values of the flow parameters
has been already
found in black hole accretion (Park 1990; Nobili, Turolla \& Zampieri 1991)
when both free--free and Compton scattering are present.
An alternative
picture for the production of hard X--rays ($\sim 100$ keV) from accreting
neutron stars has been proposed by Klu\'zniak \& Wilson (1991). In their model
matter coming from the inner edge of an accretion disk hits the stellar surface
at a shallow angle, creating a hot equatorial belt in which Compton cooling
is very efficient.

\beginsection III. DISCUSSION

The basic limitations of our approach are obviously that the
analysis is stationary and the spectrum is described just in terms of a
mean photon energy.
The fact that ``hot'' models
may exist only for high enough radiation temperatures,
$\Tg \magcir T_{crit} \simeq 10^9$K
at the outer boundary, suggests that, in order to get started,
an extra energy input, different
from that produced by the incoming beam, must be supplied,
the actual value of $\Tg$ depending on the preparation of the system.
A physically consistent scenario should be investigated in a time dependent
picture.
Furthermore temperatures are mildly relativistic
and the expression we used for the Compton
heating--cooling term is just an approximation.
Moreover, pair production--annihilation
can not be neglected  for
$l_{\infty}\magcir 10^{-2}$ when
the compactness parameter becomes $\magcir 10.$
Above $l_{\infty} \sim 0.1,$
the dynamical effects of radiation pressure and bulk motion Comptonization
become also important (on this regard see e.g. Zampieri, Turolla \& Treves 1993
and references therein) and are not included in our model.

Our present results can be summarized as follows.
At very low accretion rates (the threshold depends on $y_0$) there is a unique
solution, the ``cold'' one discovered by ZS (see figure 3). This essentially
indicates that the black body approximation at very low luminosities, as
expected for instance in isolated neutron stars accreting the interstellar
medium is reasonable
(see e.g. Ostriker, Rees \& Silk 1970; Treves \& Colpi 1991; Blaes
\& Madau 1993). The ``hot'' and ``cold'' solutions coexist for
a certain range of luminosities and the mean photon energies of the two
modes approach each other for increasing $l_\infty$.
The ``cold'' solution becomes hotter for increasing accretion rate while
the ``hot'' one softens and this behaviour is opposite to the one
exhibited by the hot, shocked solutions of Shapiro \& Salpeter.

Even in the absence of frequency--dependent
calculations it is tempting to associate
the ``cold'' solution with the spectral states observed in the
Soft X--Ray Transient source Aql X--1, either in outburst or in quiescence
(Verbunt et al. 1993).
The existence of ``hot'' solutions for intermediate luminosities with
temperatures $\sim 100$ keV suggests that a class of hard X--ray sources,
which possibly have not yet been discovered, may exist.
The transition between the ``hot'' and the ``cold'' regime, even at
luminosities
where the two solutions are rather different,
may be expected in a time--dependent scenario.
Non--stationary calculations are also needed to explore
the stability properties of the two solutions.

In conclusion  the discovery of ``hot'' solutions in
spherical accretion onto neutron stars deserves further
theoretical consideration, a program on which we are presently actively
working.

\beginsection ACKNOWLEDGMENTS

We thank an anonymous referee for some helpful comments.

\beginsection REFERENCES

\ref{Alme, M.L., \& Wilson, J.R. 1973, ApJ, 186, 1015}
\ref{Bildsten, L., Salpeter, E.E., \& Wasserman, I. 1992, ApJ, 384, 143}
\ref{Blaes, O., \& Madau, P. 1993, ApJ, 403, 690}
\ref{Klu\'zniak, W., \& Wilson, J.R. 1991, ApJ, 372, L87}
\ref{London, R.A., Taam, R.E., \& Howard, W.E. 1986, ApJ, 306, 170}
\ref{Nobili, L., Turolla, R., \& Zampieri, L. 1991, ApJ, 383, 250}
\ref{Ostriker, J.P., Rees, M.J., \& Silk, J. 1970, Astrophys. Letters, 6, 179}
\ref{Park, M--G. 1990, ApJ, 354, 64}
\ref{Park, M--G., \& Ostriker, J.P. 1989, ApJ, 347, 679}
\ref{Shapiro, S.L., \& Salpeter, E.E. 1973, ApJ, 198, 761}
\ref{Treves, A., \& Colpi, M. 1991, A\&A, 241, 107}
\ref{Verbunt, F., Belloni, T., Johnston, H.M., van der Klis, M., \&
Lewin W.H.G. 1993, A\&A submitted}
\ref{Zampieri, L., Turolla, R., \& Treves, A. 1993, ApJ, 419, 311}
\ref{Zel'dovich, Ya., \& Shakura, N. 1969, Soviet Astron.--AJ, 13, 175}
\ref{Wandel, A., Yahil, A., \& Milgrom, M. 1984, ApJ, 282, 53}

\vfill\eject
\beginsection FIGURE CAPTIONS

\refig{Figure 1.\quad {Temperature vs. column density of ``cold'' solutions
for $l_{\infty} = 7 \times 10^{-3}$ (continous line),
$l_{\infty} = 2 \times 10^{-2}$ (dashed line) and
$l_{\infty} = 7 \times 10^{-2}$ (dashed--dotted line);
here $y_0 = 20$.}
\medskip
\refig{Figure 2.\quad {Same as in figure 1 for ``hot'' solutions.}
\medskip
\refig{Figure 3.\quad {Mean energy of the
outgoing photons vs. total emitted luminosity for $y_0 = 20$;
crosses refer to ``cold'' models.
The dashed line represents the lower limit for the existence of ``hot''
solutions given by equation (13).}

\vfill\eject\bye

\bye